\documentclass[lettersize,journal]{IEEEtran}
\usepackage{amsmath,amsfonts}
\usepackage{algorithmic}
\usepackage{array}
\usepackage{textcomp}
\usepackage{stfloats}
\usepackage{url}
\usepackage{verbatim}
\usepackage{graphicx}
\usepackage{makecell}
\usepackage{subcaption}
\usepackage{caption}
\def\BibTeX{{\rm B\kern-.05em{\sc i\kern-.025em b}\kern-.08em
    T\kern-.1667em\lower.7ex\hbox{E}\kern-.125emX}}
\usepackage{balance}

\usepackage{hyperref}
\hypersetup{
	colorlinks=true,  
	linkcolor=red,    
	citecolor=blue,   
	urlcolor=black   
}

\begin{document}
\title{
New Paradigm for Unified Near-Field and Far-Field Wireless Communications

  
  }

\author{Zhaocheng~Wang,~\emph{IEEE Fellow},~Haochen~Wu, Yuanbin~Chen,~and~Liyang~Lu,~\emph{IEEE Member}

\thanks{
Zhaocheng~Wang, Haochen~Wu, and Liyang~Lu are with the Department of Electronic Engineering, Tsinghua University, Beijing 100084, China (e-mails: zcwang@tsinghua.edu.cn, wuhc23@mails.tsinghua.edu.cn, luliyang@mail.tsinghua.edu.cn). (\textit{Corresponding author: Liyang Lu.})

Yuanbin~Chen is with the School of Electrical and Electronics Engineering, Nanyang Technological University, Singapore 639798 (email: yuanbin.chen@\protect\\ntu.edu.sg).

Zhaocheng Wang and Haochen Wu contributed equally to the project and should be considered co-first authors.}
\vspace{-1em}
}

\maketitle

\begin{abstract}
Current Type~I and Type~II codebooks in fifth generation (5G) wireless communications are limited in supporting the coexistence of far-field and near-field user equipments (UEs), as they are exclusively designed for far-field scenarios. To fill this knowledge gap and encourage relevant proposals by the 3rd Generation Partnership Project (3GPP), this article provides a novel codebook to facilitate a unified paradigm for the coexistence of far-field and near-field contexts. It ensures efficient precoding for all UEs, while removing the need for the base station to identify whether one specific UE stays in either near-field or far-field regions. Additionally, our proposed codebook ensures compliance with current 3GPP standards for working flow and reference signals. Simulation results demonstrate the superior performance and versatility of our proposed codebook, validating its effectiveness in unifying near-field and far-field precoding for sixth-generation~(6G) multiple-input multiple-output (MIMO) systems.

\end{abstract}


\section{Introduction}\label{S1}

Recent years have witnessed significant advancements in sixth-generation (6G) wireless communications \cite{dai2022tcom,wu2024tcom}. Featured by higher frequency bands of Frequency Range 1 and Frequency Range 2, along with larger antenna apertures adopted by the extremely large-scale multiple-input-multiple-output (XL-MIMO) technology, future 6G wireless systems promise abundant spectrum resources and a ten-fold increase in the spectral efficiency~\cite{chen2024wcl}. However, these advancements also usher in fundamental shifts in electromagnetic (EM) wave propagation, particularly the emergence of near-field effect~\cite{wu2024tcom}.
At the heart of this phenomenon lies the distinction between far-field and near-field regions, defined by Rayleigh distance, which scales with both the antenna aperture and the carrier frequency~\cite{wu2024tcom}.
Beyond the Rayleigh distance is the far-field region, where EM waves are typically approximated as planar counterparts. In this regime, far-field steering vectors depend solely on the propagation direction and are conveniently represented by discrete Fourier transform (DFT) vectors. By contrast, the near-field region lies within the Rayleigh distance, where the planar-wavefront assumption breaks down, necessitating spherical wave modeling. Hereby, steering vectors are influenced by both the propagation direction and the distance, leading to new challenges for near-field regimes. These dual-region dynamics mark a paradigm shift in 6G MIMO design, requiring novel models and approaches.


\begin{table*}[tp!]
	  \caption{Recent efforts toward near-field communications}\label{table1}
	  \centering
  \begin{tabular}{|c|c|c|c|}
    \hline
    Reference           & Channel model                        & Research issues                     & Main contributions                                                \\
    \hline \hline
    \cite{dai2022tcom}  & Near-field                           & Channel estimation                  & \makecell{Propose the polar-domain codebook for                   \\ near-field channels with uniform linear arrays} \\
    \hline
    \cite{wu2024tcom}   & Near-field                           & Channel estimation                  & \makecell{Leverage side-information from lower frequency        \\ band to assist near-field channel estimation} \\
    \hline
    \cite{dai2023jsac}  & Near-field                           & Codebook-based hybrid precoding     & \makecell{Design the precoding method for                         \\ near-field channels with uniform planar arrays} \\
    \hline
    \cite{yue2024twc}   & Hybrid-field                         & Channel estimation                  & \makecell{Estimate and eliminate power diffusion when using joint \\ angular-polar transformation to express channel sparsity} \\
    \hline
    \cite{xi2024wcl}    & Hybrid-field                         & Channel estimation                  & \makecell{Propose gridless channel estimation based on 
    \\ compressive sensing and fractional DFT} \\
    \hline
    \cite{zheng2024wcl} & \makecell{Coexisting near/far-field} & Optimization-based hybrid precoding & \makecell{Propose joint optimization of UE selection, hybrid            \\ precoder and rate allocation without codebooks} \\
    \hline
  \end{tabular}
  \label{tab:addlabel}
  \captionsetup{justification=raggedright, singlelinecheck=false}
\end{table*}

{Extensive research efforts from both academia and industry have been dedicated to addressing the near-field issues. Table~\ref{table1} provides a summary of key contributions to near-field investigations from academic institutions, where most studies focus primarily on specific scenarios, such as those limited to the near-field context~\cite{wu2024tcom,dai2023jsac}. {Existing hybrid-field scenarios usually consider the single user equipment (UE) case, where the wireless channel consists of both near-field and far-field components, neglecting the more realistic and prevalent multiple UE scenarios \cite{yue2024twc,xi2024wcl}.} Even in coexisting near/far-field scenarios involving the UEs located in either far-field or near-field regions, the codebook design often relies on the prior knowledge of each UE's position, that is, whether it stays in either far-field or near-field regions~\cite{zheng2024wcl}. The acquisition of such information is sometimes unrealistic or requires additional pilots, which introduces unnecessary overhead and limits implementations in practical systems.

From an industrial perspective, the 3rd Generation Partnership Project (3GPP) has established a working item (WI) dedicated to investigating near-field channel characteristics. This initiative focuses on extending the channel model specified in TR~38.901~\cite{TR-38901}, particularly for the 7-24 GHz frequency range, to address the spatial non-stationarity inherent in near-field propagation~\cite{3GPP-2023}. At the present time, consensus has been reached on several aspects of near-field standardization in the most recent 3GPP technical specification group radio access network (TSG RAN) meeting \#105~\cite{3GPP-2024}, which includes defining near-field region, specifying near-field channel parameters, and modeling near-field channels across diverse deployment scenarios such as urban macro (UMa), urban micro (UMi), indoor office, and indoor factory environments. Furthermore, 3GPP RAN1 group is striving to design a unified framework that explicitly reflects the characteristics of near-field propagation alongside the existing properties of far-field propagation within the existing TR~38.901 stochastic model. However, the remaining open issues accounting for the spatial non-stationarity in near-field propagation while ensuring consistency between near-field and far-field models need to be solved.

Given collective efforts from both academia and industry, we deem that constructing a unified framework for accommodating near-field and far-field ingredients is an imperative objective, which need to be diligently pursued and finally realized.
Beyond the open issues identified by 3GPP, a more straightforward challenge in coexistence near-field/far-field scenarios lies in multi-UE precoding at the base station (BS). To serve multiple UEs simultaneously, inaccurate acquisition of channel state information (CSI) for one specific UE may severely degrade the precoding performance from all other UEs, creating a ``domino effect'', which underscores the critical dependency of multi-UE precoding on the accurate CSI acquisition for each UE. However, current Type~I and Type~II codebooks in 5G wireless systems,  tailored exclusively for far-field regimes, are inadequate for the coexistence of near-field/far-field scenarios. To circumvent this issue, this article explores the bespoke codebook design for accommodating both far-field and near-field regimes.
In particular, we commence by reviewing Type~I and Type~II codebooks defined in 5G wireless systems, highlighting their limitations when employed for far-field and near-field coexisting scenarios. After that, a novel codebook customized for supporting the coexistence of far-field and near-field precoding is introduced. Simulation results verify the accuracy of CSI acquisition in the wavenumber domain and demonstrate the significant spectral efficiency improvement in multi-UE precoding scenarios, regardless of whether users are located in either far-field or near-field regions. Finally, several future research directions are discussed.


}



\section{Overview of Type~I and Type~II Codebooks}\label{S2}

\begin{figure*}
	\begin{center}
		\includegraphics[width=1\textwidth]{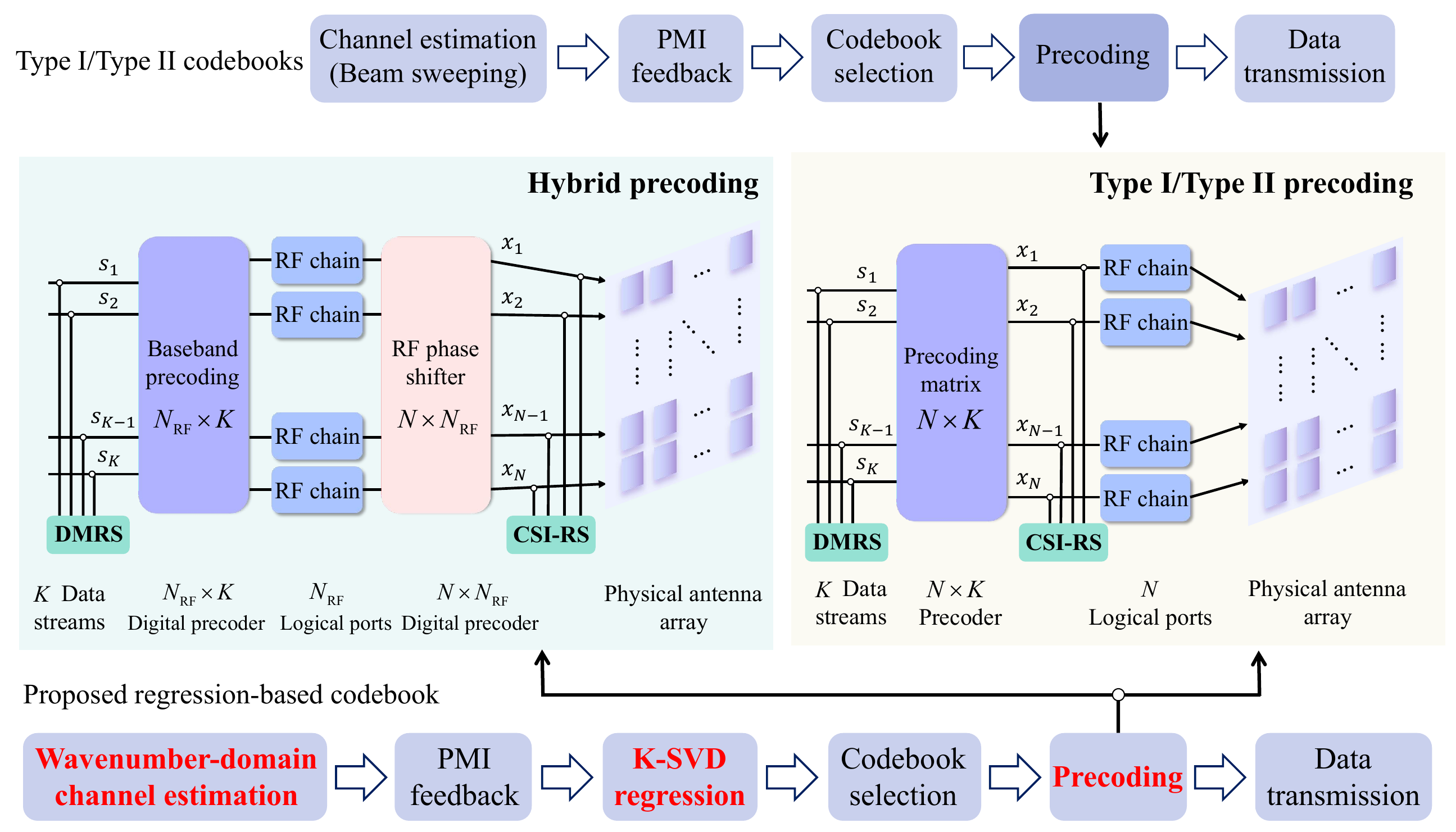}
	\end{center}
	\captionsetup{justification=raggedright, singlelinecheck=false}
	\caption{{Illustration of generation process and precoding architecture of Type~I/Type~II and our proposed codebooks. 
	Firstly, the right part illustrates Type~I/Type~II precoding. $K$ data streams and DMRS are multiplexed and sent to the precoding matrix, which assigns $K$ data streams to $N$ RF chains. 
	The output of precoding matrix, along with CSI-RS, 
	are beamformed to $N$ logical ports and then mapped into the corresponding physical antennas. 
	Secondly, the hybrid precoding on the left part follows the architecture presented in \cite{alkhateeb2015twc}, 
	where $K$ data streams together with DMRS are delivered to $N_{RF}$ RF chains adopting a $N_{RF}\times K$ baseband precoding matrix.
	Then, an analog RF phase shifter array, represented by a $N \times N_{RF}$ matrix, maps the corresponding RF chains into $N$ physical antennas.}
	}
	\label{fig3_proposed}
\end{figure*}

{

In this section, we commence by reviewing the mechanisms of Type~I and Type~II codebooks in 5G wireless systems, as well as the precoding architectures that support their implementations at the BS. We then extend our discussion to the coexistence scenarios, highlighting the limitations of these codebooks in such contexts.

\subsection{Type~I and Type~II Codebooks}
The right part of Fig.~\ref{fig3_proposed} illustrates the workflow of current Type~I and Type~II codebooks, which are outlined as follows.
\begin{itemize}
\item Channel estimation (beam sweeping): In 5G wireless systems, reference signals (RS) are employed for channel estimation, including synchronization signal blocks (SSB), demodulation reference signals (DMRS), and channel state information reference signals (CSI-RS).

\item PMI feedback: The precoding matrix indicator (PMI) and the rank indicator for multiple data streams are incorporated into the CSI report. 
In other words, the PMI carries basis indices related to the primary CSI intended for the BS.

\item Codebook selection: There are two predefined formats for PMI in 5G wireless systems: Type~I and Type~II. Type~I PMI offers the advantage of low overhead and computational simplicity, making it easier for each UE to calculate. In contrast, Type~II PMI provides greater flexibility, making it well-suited for multi-UE MIMO applications. However, this increased flexibility comes at the cost of higher computational complexity and overhead. Type~I and Type~II PMIs are also known as codebooks.

\item Precoding: The multiple data streams (also referred to as data layers) are each associated with DMRS, which are then precoded over logical antenna ports and mapped to the physical antennas for actual transmission. 
\end{itemize}
Regarding the codebook selection, i.e., Type~I or Type~II, we elaborate upon their subtle differences as follows.

}

\subsubsection{Type~I Codebook}
Type~I codebook is based on quantizing the beam directions according to the antenna steering vector.
In far-field scenarios, the DFT matrix and its oversampling counterparts are commonly adopted as the codebook, due to their strong alignment with the far-field channel response.
Firstly, BS obtains the CSI through channel estimation,
and then selects one codeword with highest correlation for precoding.
However, the Type~I codebook precoding scheme has two main shortcomings.  On the one hand, only the strongest path is chosen for precoding, whereas the UEs usually experience multi-path wireless channels.
This not only limits the maximum achievable spectral efficiency, but also causes a sharp drop in precoding performance if multiple UEs share the same strongest path.
On the other hand, the Type~I codebook requires perfect CSI, resulting in unacceptable feedback overhead when the number of UEs becomes large.

\subsubsection{Type II Codebook}
{
Type~II codebook, in stark contrast to its Type~I counterpart, performs beam sweeping with reduced overhead for CSI acquisition, rather than depending exclusively on the channel estimation.} To be specific, Type~II codebook explores the spatial sparsity potential in the far-field channels, allowing the entire channel matrix to be well represented by several significant paths.
When Type~II codebook is adopted, BS sequentially applies each codeword and transmits pilot signals that undergo the precoding process accordingly. After that, UEs then report the received values to the BS, delivering a coarse estimation of the overall CSI. Upon receiving feedback from UEs, the BS selects a few codewords with the highest estimate magnitudes, which corresponds to several significant paths between the BS and its serving UEs.
Finally, the precoding process adopts a linear combination of the selected codewords, utilizing those significant paths for data transmission to each specific UE. This strategy overcomes the limitation of Type~I codebook, where some UEs might be constrained by sharing the same strongest path.

{
\subsection{Challenges}
Existing codebook schemes, specifically Type~I and Type~II, are fundamentally based on the assumption of far-field planar wave propagation. This dependency on DFT-based processing leads to significant power diffusion or leakage within the angular power distribution, as presented in~\cite{chen-cm,guo2024icc}. Such power diffusion results in significant inaccuracies in channel estimation because the power of significant path is not accurately captured, subsequently degrading the precoding performance. This issue becomes particularly critical in multi-UE scenarios. An erroneous CSI acquisition of one specific UE located at near-field region adversely impacts the precoding performance for all other users within the cell, resulting in a ``domino effect''. 
While recent studies have proposed near-field channel estimation methods and codebook design schemes~\cite{dai2022tcom,wu2024tcom,dai2023jsac}, these polar-domain techniques require non-uniform distance sampling based on specific angular directions. This introduces two major issues: i) BS has to receive reports from its serving UEs indicating whether they are in the near-field or far-field region, leading to high pilot overhead; ii) non-uniform distance sampling faces the curse of calculation dimensionality, rendering these methods less practical for implementations.
In pursuit of practicability, 3GPP is actively working towards developing a near-field channel model compatible with the existing far-field channel model defined in TR~38.901, which aims to integrate near-field ingredients and establishs a unified paradigm, which could accommodate both far-field and near-field communications. Resolving this issue possesses high priority for current standards development efforts, underscoring an urgent need to bridge the gap between near-field and far-field communication scenarios.

}


\section{Codebook Design for Unified Near-Field and Far-Field Communications}

{
Geared towards challenges of current 5G codebooks in addressing the coexistence of far-field and near-field scenarios, this section devises a novel codebook methodology, which is illustrated in the bottom part of Fig.~\ref{fig3_proposed}. In contrast to the diagram depicted in the top part of Fig.~\ref{fig3_proposed}, our design maintains compatibility with existing 5G standardization while emphasizing three key distinctions including wavenumber-domain channel estimation, K-singular value decomposition (K-SVD) regression and precoding.


}


\subsection{Wavenumber-Domain Channel Estimation}\label{S31}

Since the precoding performance is significantly dependent upon the accuracy of CSI acquisition, inaccurate channel estimation may significantly degrade the performance of spectral efficiency~\cite{chen-cm}. This issue is particularly pronounced in multi-UE scenarios, where erroneous CSI feedback from one user may adversely impact the spectral efficiency of other users within the same cell, resulting in a ``domino effect''. To eliminate the inaccuracy of near-field CSI acquisition, our proposed codebook scheme firstly necessitates channel estimation in the wavenumber domain. More specifically, by leveraging the Fourier harmonic expansion, both far-field and near-field channel responses can be represented as the superposition of a series of planar waves characterized by different wavenumbers~\cite{pizzo2022twc,chen2024wcl}. This differs from both Fraunhofer and Fresnel approximations usually employed in far-field and near-field scenarios. 
Based on the wavenumber-domain channel representation, efficient channel estimation algorithms are proposed accordingly [12], [13], in order to explore the potential sparsity determined by the environmental scatterers in the wireless channels.
To be specific, by modeling all signal processing modules as the measurement matrix,
the wavenumber-domain channel estimation could be transformed into a sparse recovery problem based on the wavenumber-domain sparse representation,
which can be handled by various compressive sensing algorithms, i.e. orthogonal matching pursuit (OMP) or Markov random field (MRF).

\begin{figure}[t]
	\begin{center}
		\includegraphics[width=0.47\textwidth]{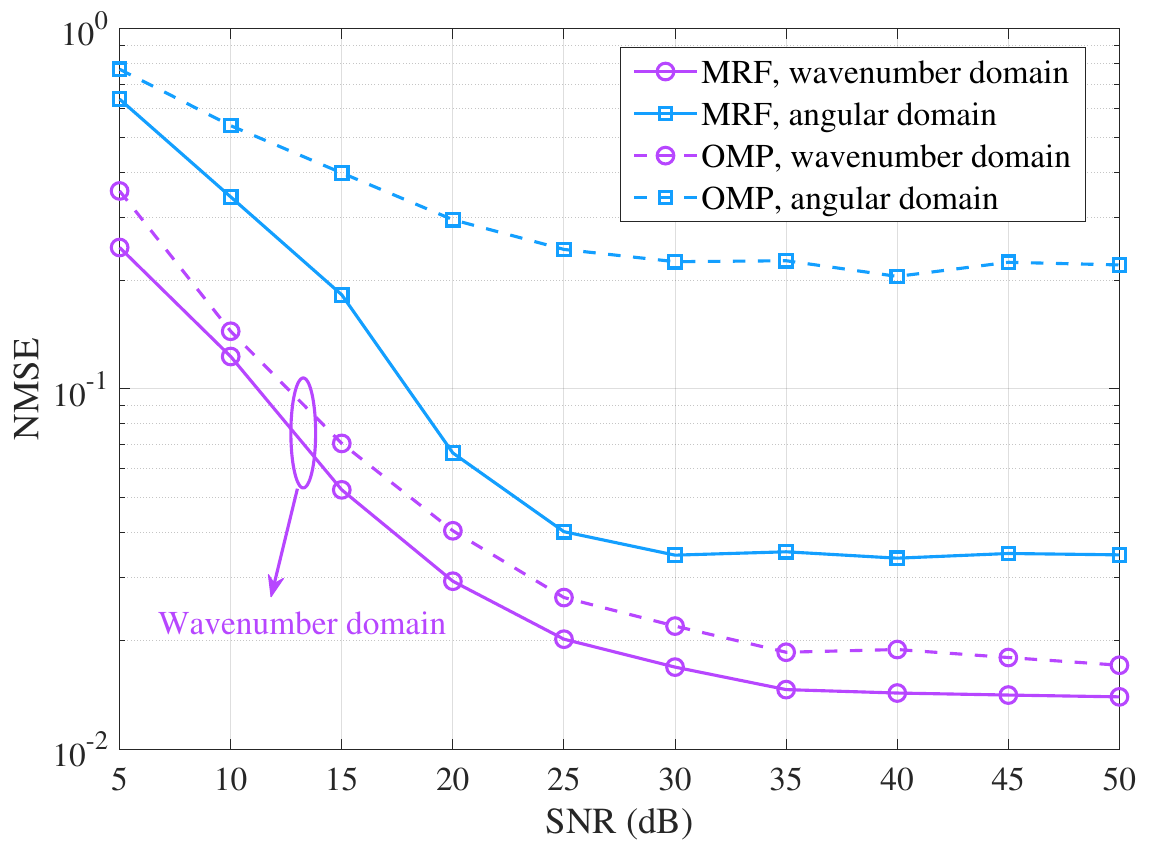}
	\end{center}
	\caption{Performance comparison of different channel estimations in both angular and wavenumber domains.}
	\label{NMSEvsSNR}
\end{figure}

Fig.~\ref{NMSEvsSNR} illustrates the channel estimation performance achieved by the MRF-based algorithm~\cite{guo2024icc} and the OMP~\cite{wu2024tcom}. The performance is valued in terms of the normalized mean squared error (NMSE). Usually, MRF is implemented in the wavenumber domain, while OMP operated in the angular domain. It is observed from Fig.~\ref{NMSEvsSNR} that regardless of whether OMP or MRF algorithm is employed, channel estimation in the wavenumber domain consistently outperforms its counterpart in the angular domain. This superior performance is attributed to the wavenumber-domain modeling, which effectively addresses the power diffusion problem inherent in coexisting near-field and far-field scenarios through Fourier harmonic decomposition.
Moreover, the MRF channel estimation could capture the elliptical clustering features of non-zero entries in the wavenumber domain,
leading to the improved recovery performance \cite{guo2024icc}.

\subsection{K-SVD Regression} \label{S32}

\begin{figure*}[t]
	\begin{center}
		\includegraphics[width=0.76\textwidth]{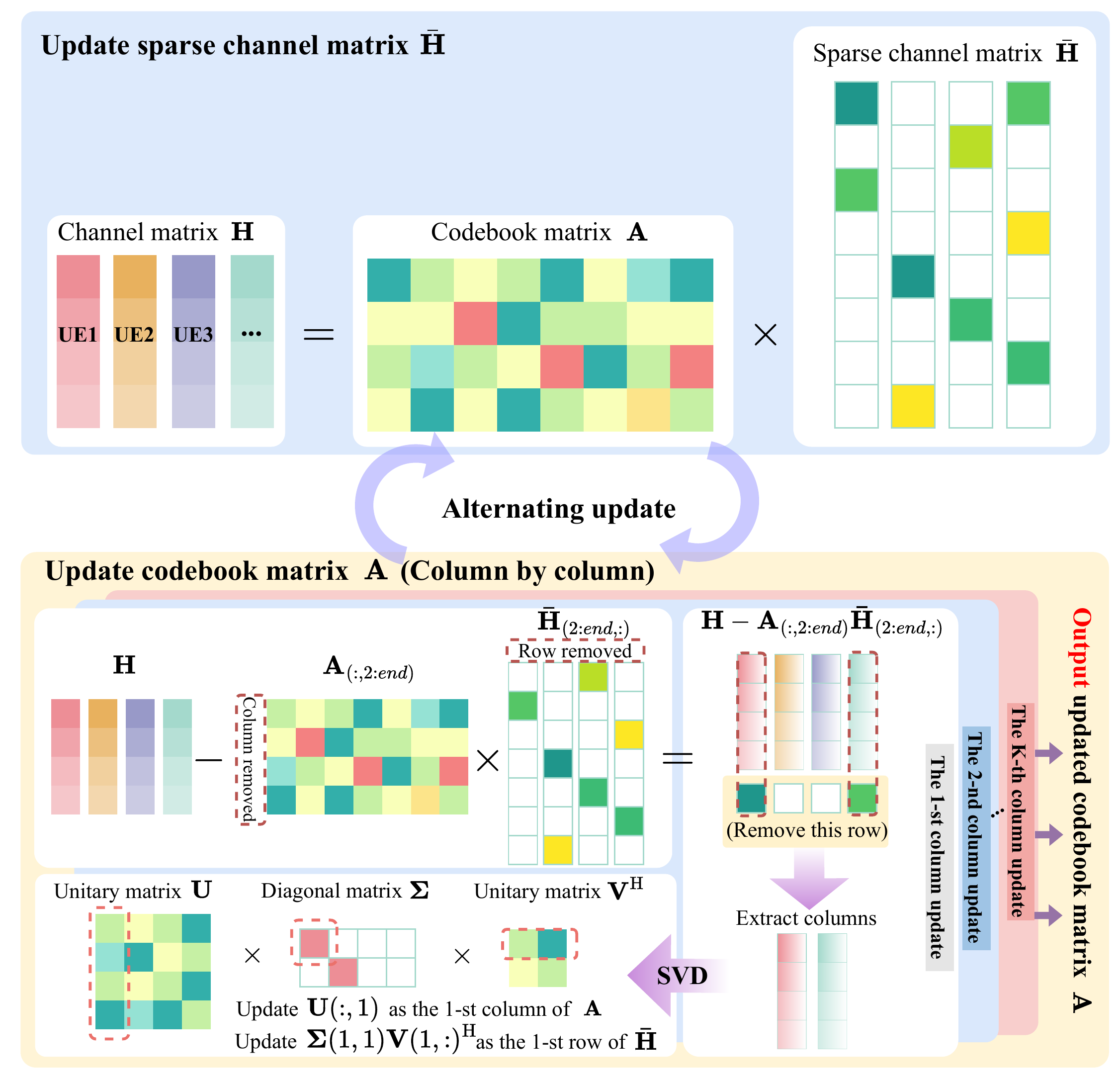}
	\end{center}
	\captionsetup{justification=raggedright, singlelinecheck=false}
	\caption{Illustration of the proposed codebook generation.
	K-SVD algorithm is leveraged to obtain our proposed codebook,
	where the codebook matrix $\mathbf{A}$ and the corresponding sparse channel matrix $\bar{\mathbf{H}}$ 
	are alternatively optimized using the compressive sensing algorithm and the SVD decomposition technique,
	respectively.
	}
	\label{KSVD}
\end{figure*}

After obtaining the PMI feedback information at the BS, it is crucial to design new codebook in place of predefined Type~I and Type~II codebooks. For this purpose, we firstly employ the K-SVD algorithm as the foundation of our codebook design. Specifically, K-SVD is a well-established dictionary learning method that decomposes input data samples, i.e., the original channel matrix $\mathbf{H}$, into a codebook matrix $\mathbf{A}$ and a sparse channel matrix denoted by $\bar {\mathbf{H}}$. The objective is to minimize the NMSE between the original channel matrix $\mathbf{H}$, and its approximation ${\mathbf{A}} {\bar {\mathbf{H}}}$. The algorithm operates through an iterative optimization process, as illustrated in Fig.~\ref{KSVD}.
\begin{itemize}
\item Update sparse channel matrix $\bar{\mathbf{H}}$: Given a fixed codebook matrix $\mathbf{A}$, the K-SVD algorithm tries to solve the sparse channel matirx $\bar{\mathbf{H}}$. It involves addressing multiple sparse recovery problems simultaneously, which can be efficiently handled using compressive sensing techniques, such as OMP~\cite{wu2024tcom}.

\item Updating codebook matrix $\mathbf{A}$: Given a fixed $\bar{\mathbf{H}}$, the algorithm updates the codebook matrix $\mathbf{A}$ one column at a time. For each update, SVD contributes to a column of the codebook matrix $\mathbf{A}$, ensuring that the codebook matrix $\mathbf{A}$ is optimally tuned to represent the input channel realizations accurately.
\end{itemize}
This alternating optimization continues until the NMSE falls below a predefined threshold or the maximum number of iterations is reached. The output is the designed codebook matrix $\mathbf{A}$ and a sparse channel matrix $\bar{\mathbf{H}}$. 
After obtaining the desired codebook by K-SVD, the codebook selection process is the same as the conventional procedure defined in 5G standard. Specifically, BS transmits CSI-RS that carries each codeword within the codebook. 
Upon receiving these pilots, UE reports the indices of the most related codewords it identifies.
In practical scenarios, acquiring accurate $\mathbf{H}$ periodically
may lead to high computational complexity and signaling overhead.
Efficient approaches, i.e., offline learning, might be considered to reduce the implementation cost,
where the BS only obtains the codebook $\mathbf{A}$ based on an initial CSI dataset.
$\mathbf{A}$ remains unchanged until a substantial decline in spectral efficiency is observed,
requiring another K-SVD optimization of $\mathbf{A}$ and sparse channel matrix $\mathbf{\bar{H}}$.

\subsection{Precoding}



In the proceding process, either of these two precoding architectures illustrated in Fig.~\ref{fig3_proposed} can be employed.
However, the hardware limitations of phase shifters in analog precoding impose a constant modulus constraint on the codewords, requiring that all entries should have the same modulus in the codebook matrix. Therefore, after obtaining the codebook matrix $\mathbf{A}$ via K-SVD, appropriate post-processing is required. To be specific, the projection operation is conducted on $\mathbf{A}$, dividing each entry by its own modulus~\cite{zheng2024wcl}. 
Despite slightly eroding the sparsity of $\bar{\mathbf{H}}$, it has an acceptable impact on the precoding performance. This trade-off can be comparatively negligible when weighed against larger challenges, such as the misalignment of beam directions and the inherent discrepancies in channel model assumptions observed in Type~I and Type~II codebooks.

\begin{figure}[h!]
	\centering
	\begin{subfigure}[b]{0.45\textwidth}
		\centering
		\includegraphics[width=\textwidth]{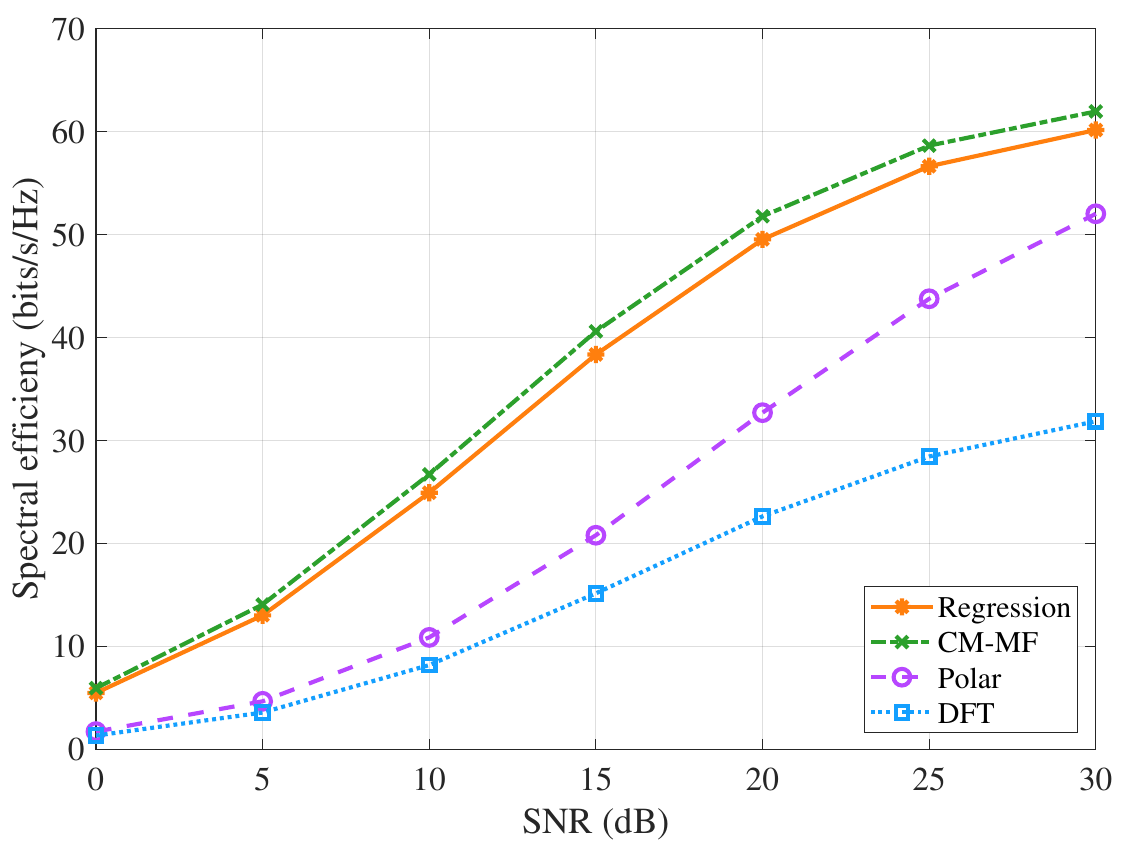}
		\caption{}
		\label{fig_4a}
	\end{subfigure}
	\hfill
	\begin{subfigure}[b]{0.45\textwidth}
		\centering
		\includegraphics[width=\textwidth]{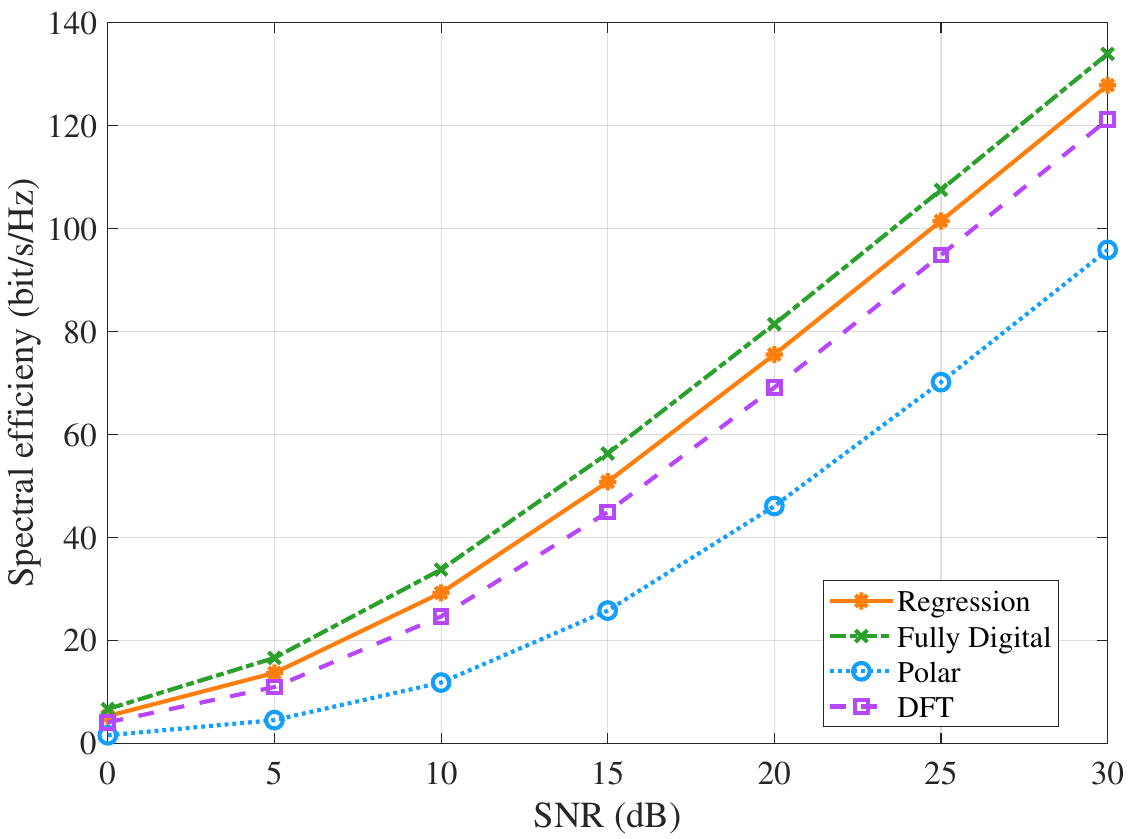}
		\caption{} 
		\label{fig_4b}
	\end{subfigure}
	\caption{(a) Spectral efficiency comparison with beam sweeping from DFT, Polar and our proposed regression methodology; (b) Spectral efficiency comparison with hybrid precoding from DFT, Polar and our proposed regression methodology.} 
	\label{fig4}
\end{figure}

\begin{figure}[h!]
	\centering
	\begin{subfigure}[b]{0.42\textwidth}
		\centering
		\includegraphics[width=\textwidth]{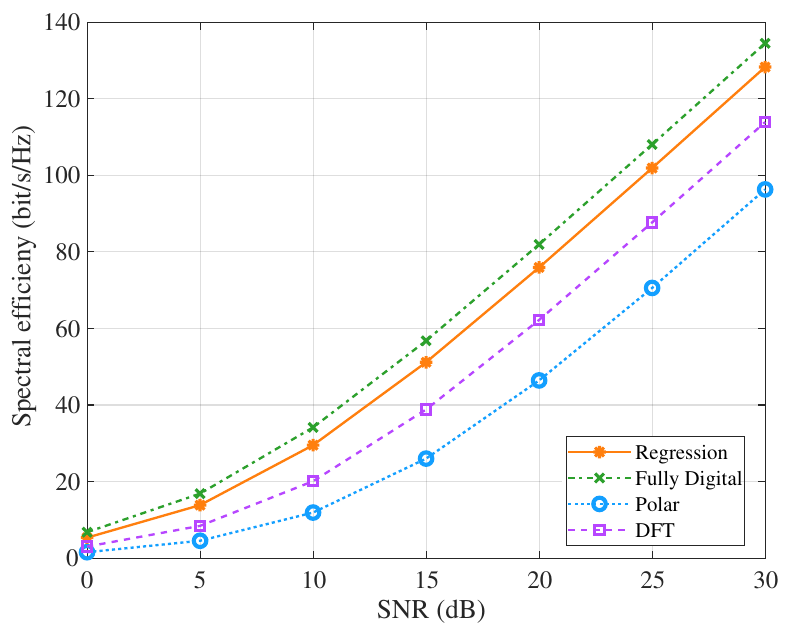}
		\caption{}
		\label{fig_5a}
	\end{subfigure}
	\hfill
	\begin{subfigure}[b]{0.42\textwidth}
		\centering
		\includegraphics[width=\textwidth]{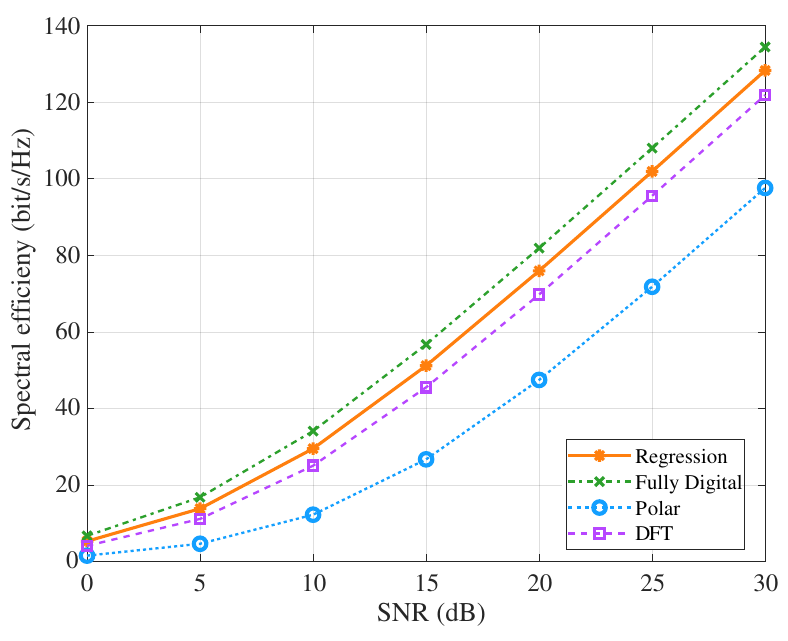}
		\caption{} 
		\label{fig_5b}
	\end{subfigure}
	\caption{(a) Spectral efficiency comparison using hybrid precoding when UEs are all located in the near-field region; (b) Spectral efficiency comparison using hybrid precoding when UEs are all located in the far-field region.} 
	\label{fig5}
\end{figure}
{
To verify the effectiveness of our proposed codebook scheme, Fig.~\ref{fig_4a} employs the beam sweeping approach (i.e., Type~I/Type~II precoding illustrated in the right part of Fig.~\ref{fig3_proposed}), while Figs.~\ref{fig_4b}, \ref{fig_5a}, and \ref{fig_5b} utilizes the hybrid precoding architecture~\cite{alkhateeb2015twc}. In all subfigures, ``Regression" refers to our proposed codebook, ``Polar" corresponds to the scheme presented in~\cite{dai2022tcom,dai2023jsac}, and ``DFT" represents the standard Type~II codebook specified in 5G NR. Additionally, Fig.~\ref{fig_4a} includes the constant modulus matched filter (CM-MF) scheme,
where the transpose of the perfect CSI matrix is leveraged to be the codebook, 
acting as a benchmark for analog beamforming, while Figs.~\ref{fig_4b}, \ref{fig_5a}, and \ref{fig_5b} incorporate fully digital precoding to establish an upper performance bound.
We consider one BS equipped a uniform planar array (UPA) having $ 32 \times 32=1024 $ antenna elements, serving 16 UEs. 
In Figs. 4a and 4b, 4 of these UEs are distributed randomly in the near-field region and 12 are randomly located in the far-field region.
On the other hand, all 16 users are randomly distributed solely within the near-field region in Fig. 5a, and Fig. 5b depicts a scenario 
where all users are randomly distributed in the far-field region.	
We follow the channel model prescribed in 3GPP~TR~38.901~\cite{TR-38901}, where each user's channel consists of 4 clusters, with each cluster containing 5 rays,
and $500$ simulations are done to obtain the average performance.

In Figs.~\ref{fig_4a} and \ref{fig_4b}, our proposed regression-based codebook outperforms its DFT and polar counterparts, and even demonstrates a spectral efficiency close to the theoretical upper bound achieved by fully digital precoding. Similar results can be obtained in Figs.~\ref{fig_5a} and \ref{fig_5b}. It is evident that K-SVD algorithm excels at modeling and approximating the steering vectors of the original channel matrix  $\mathbf{H}$, which allows our proposed methodology to dynamically adjust the weights of its codewords based on the specific characteristics of the channel. Unlike traditional codebooks with fixed codewords and directions, the regression-based codebook assigns higher correlations to codewords that align with steering vectors exhibiting stronger gains in $\mathbf{H}$. Additionally, accurate CSI acquisition in the wavenumber domain greatly ensures the performance of our proposed codebook, which can be reflected by the poor performance achieved by Polar and DFT schemes. 
Therefore, the combination of accurate CSI acquisition in the wavenumber domain and the great robustness of K-SVD could provide a double assurance for the precoding performance across diverse wireless scenarios, effectively overcoming the limitations posed by traditional Type~I and Type~II codebooks.
Furthermore, when the beam sweeping approach is leveraged, Polar exhibits less power diffusion and more accurate codeword selection than DFT due to near-field effects, 
thus achieving better spectral efficiency.
However, when the hybrid precoding architecture is considered, the analog RF phase shifter can eliminate the power diffusion effects induced by the use of the DFT codebook.

}


\section{Future Prospects}
\label{S5}
This section provides several future research directions in terms of unified far-field and near-field wireless communications, with an outlook on anticipated challenges in the road to 3GPP 6G standardization.

\subsection{Refining RS for Unified Paradigm}

Current 3GPP standards primarily address far-field scenarios, with RS specifically designed for planar-wave channel models. However, the unified far-field and near-field paradigm necessitates the redefinition and redesign of RS signaling. Future research may focus on developing new RS signaling (e.g., DMRS, CSI-RS) that are compatible with both far-field and near-field scenarios and at the same time enable accurate and reliable feedback, ensuring seamless integration within existing standards such as TR 38.901. 

\subsection{Wavenumber-Domain Multiple Access}
Wavenumber-domain modeling accounts for evanescent waves, which are particularly significant in near-field communications but are typically neglected in far-field scenarios. These evanescent waves introduce additional degrees of freedom, thereby contributing to wavenumber-domain multiple access (WDMA). Recent studies have confirmed that WDMA can significantly enhance system capacity~\cite{chen-GCw24}, although addressing wavenumber-domain interference remains a crucial challenge in multi-UE scenarios. Furthermore, WDMA is being considered as a promising candidate for next-generation multiple access technologies, offering the potential to meet the increasing demands of future wireless systems.

\subsection{Codebook Extendability with Continuous Aperture Array}
The concept of continuous aperture antennas is emerging, where the antenna array forms a continuous surface, such as holographic MIMO~\cite{guo2024icc}. In such systems, accurately modeling near-field effects is crucial.
Transitioning from discrete to continuous apertures requires significant modifications to fundamental channel models. Future work should explore the extension of existing wavenumber-domain and regression-based codebook to accommodate continuous aperture antennas. This inevitably involves integrating electromagnetic field theory and information theory. Such advancements will enable more accurate and efficient channel representations, enhancing system performance in future 6G MIMO technologies.

\subsection{Integrated Sensing and Communication (ISAC) with Unified Paradigm}
ISAC has also been regarded as a promising technology for future communication networks, enabling location/motion detection of surroundings objects.
A specific beamforming pattern is generated in current multi-beam ISAC schemes,
where time-invariant beams are used for communication, and other time-variant beams are allocated for sensing tasks.
Under such circumstances, near-field effects will also bring critical challenges to ISAC scenarios, since the signals from multiple mainlobes will leak into sidelobes due to power diffusion,
deteriorating both communication and sensing performance.
Therefore, considering ISAC problems in the unified near/far-field communication model, where both near-field and far-field communication are unified, is neccessary.
Future research direction may include the new ISAC channel model with near-field effects, ISAC beamforming algorithms compatible with coexisting near/far-field scenarios, and etc.

\section{Conclusion}
This article presents a novel paradigm for unified near-field and far-field communications. 
The conventional channel estimation and precoding frameworks in 3GPP standards and their challenges are firstly briefed. After that, the system model along with several key challenges, are discussed within the context of unified near/far-field wireless communications. Specifically, to establish a practical foundation for novel paradigm and realize the seamless integration of near-field and far-field communications, we introduce the wavenumber-domain channel estimation and K-SVD precoding methodology, which are proved to outperform existing benchmarks through simulation results. Finally, the future research directions are outlined within the realm of unified near/far-field communications.

\bibliographystyle{IEEEtran}
\bibliography{ref_manuscript}

\vspace {1cm}

\noindent\textbf{Zhaocheng Wang} [F] received his B.S., M.S., and Ph.D. degrees from
Tsinghua University in 1991, 1993, and 1996, respectively. From 1996 to
1997, he was a Post Doctoral Fellow with Nanyang Technological University,
Singapore. From 1997 to 2009, he was a Research Engineer/Senior Engineer
with OKI Techno Centre Pte. Ltd., Singapore. From 1999 to 2009, he was a
Senior Engineer/Principal Engineer with Sony Deutschland GmbH, Germany.
Since 2009, he has been a Professor with Department of Electronic Engineering, Tsinghua University. He was a recipient of IEEE Scott Helt Memorial
Award, IET Premium Award, IEEE ComSoc Asia-Pacific Outstanding Paper
Award and IEEE ComSoc Leonard G. Abraham Prize.

\vspace {0.5cm}

\noindent\textbf{Haochen Wu}
received the B.S. degree (Hons.) from the Department of Electronic Engineering, Tsinghua University, Beijing, China, in 2023, where he is currently pursuing the Ph.D. degree.
His research interests include wireless communications, signal processing and compressive sensing.

\vspace {0.5cm}

\noindent\textbf{Yuanbin Chen} received the Ph.D. degree (Hons.) in information and communication systems from Beijing University of Posts and Telecommunications, Beijing, China, in 2024. He is currently a Post-Doctoral Research Fellow with the School of Electrical and Electronic
Engineering, Nanyang Technological University,
Singapore. His research interests include future 6G MIMO technologies and quantum sensing.

\vspace {0.5cm}

\noindent\textbf{Liyang Lu} [M] received his Ph.D. degree from Beijing University of Posts and Telecommunications in 2022, China. He is currently a Post Doctoral Fellow with Tsinghua University. His research includes compressed sensing, cognitive radios, and MIMO communications.

\end{document}